\documentclass{elsart} 
\usepackage{graphics}
\usepackage{color}
\usepackage{epsfig}
\usepackage{times}
\usepackage{amsfonts}

\begin{document}

\begin{frontmatter}

\title{Robust ``trapping states'' in the motion of a
  trapped ion}

\author[Emmy]{S. Wallentowitz},
\author[Rostock]{W. Vogel} and
\author[Hamburg]{P.E. Toschek}

\address[Emmy]{Emmy-Noether Nachwuchsgruppe ``Kollektive
  Quantenmessung und R\"uckkopplung an Atomen und Molek\"ulen'',
  Fachbereich Physik, Universit\"at Rostock, Universit\"atsplatz 3,
  D-18051 Rostock, Germany}

\address[Rostock]{Arbeitsgruppe Quantenoptik, Fachbereich Physik,
  Universit\"at Rostock, \\ Universit\"atsplatz 3, D-18051 Rostock,
  Germany}

\address[Hamburg]{Institut f\"ur Laser-Physik, Universit\"at Hamburg,
  Jungiusstrasse 9, D-20355 Hamburg, Germany}

\begin{abstract}
  A novel robust mechanism for the generation of ``trapping states''
  is shown to exist in the coupling of a two-level system with an
  oscillator, which is based on nonlinearities in the laser-induced
  vibronic coupling.  This mechanism is exemplified with an ion
  confined in the potential well of a trap, where the nonlinearities
  are due to Franck--Condon type overlap integrals of the laser waves
  with the ionic centre-of-mass wavefunction. In contrast to the
  coherent trapping mechanism known from micro-maser theory, this
  mechanism works also in an incoherent regime operated by noisy
  lasers and is therefore much more robust against external
  decoherence effects. These features favour the incoherent regime, in
  particular for the preparation of highly excited trapping states.
\end{abstract}

\date{}

\end{frontmatter}

\section{Introduction} \label{sec:1}
Micro-maser theory predicts a coherent regime where so-called trapping
states are created in the microwave cavity field, by injection of
electronically prepared Rydberg atoms with well-defined
velocities~\cite{filipowicz}. They are stable quantum states of the
cavity field being unchanged by further interactions with injected
atoms. The trapping states are photon-number states that represent a
discrete, fixed number of cavity photons~\cite{yoo}. The preparation
of trapping states with a pre-described photon number is performed by
adjusting the atom-field interaction time, i.e., the velocity of the
injected Rydberg atoms, to certain values.  This mechanism crucially
depends on a narrow velocity distribution of the atoms and is also
very sensitive to any decoherence effects, such as cavity decay and
thermalization via background radiation, or thermal motion of the
cavity mirrors.  Another limitation is the occurrence of (rare)
two-atom events, that destroy the trapping states.  This sensitivity
is due to the fact, that the trapping mechanism relies on the coherent
features of the Jaynes--Cummings dynamics~\cite{jaynes-cummings}.
Therefore, initially not a perfect photon-number state but
sub-Poissonian photon statistics have been realised~\cite{rempe}.
Later on, signatures of trapping states with small photon number could
be successfully demonstrated~\cite{weidinger}. When preparing trapping
states of relatively large photon numbers, however, the above
mentioned detrimental influence of decoherence poses an essential
problem for the application of the coherent trapping mechanism.

Alternatively, a coherent Jaynes--Cummings type trapping mechanism can
also be realized in the motion of a single trapped
ion~\cite{toschek1,blockley,blatt,vaser}.  Here the harmonic vibration
of the trapped ion plays the role of the micro-maser cavity field, and
the interaction of the injected Rydberg atoms with the micro-maser
field is implemented by a combination of optical pumping and optical
excitation of a vibrational sideband. As in the case of the
micro-maser, this trapping mechanism for the vibration of a trapped
ion also depends crucially on a coherent Jaynes--Cummings
dynamics~\cite{blockley}.  For a realistic experimental situation with
phase-fluctuating laser fields obviously the coherent trapping
mechanism is disturbed. In that case it has been shown that a
sub-Poissonian, binomial statistics emerges for the vibrational
quantum number of the trapped ion, that reveals a noise level being
half of the classical shot noise~\cite{vaser}.

The crucial point for realizing Jaynes--Cummings type trapping states
in the motion of a trapped ion seems to be the need of a coherent time
evolution without any decoherence effects.  If one could use
incoherent dynamics to prepare the desired states, all the
above-mentioned problems could be avoided. Thus the question arises
how to realize an incoherent dynamics that also gives rise to
appropriate trapping-state conditions.

In this paper we propose a novel mechanism for the generation of
``trapping states'' in the vibration of a trapped ion that is based on
nonlinearities occurring in the laser-induced vibronic
interaction~\cite{njcm}. Dephasing effects emerging for example from
phase-fluctuating laser fields or spontaneous emissions do not disturb
this mechanism. Therefore, this incoherent regime is realistic and is
also experimentally more feasible than the coherent dynamics needed
for micro-maser-type trapping states. Moreover, the incoherent
trapping mechanism may be less sensitive to other types of externally
induced decoherence.

In Sec.~\ref{sec:2} the laser-excitation scheme and the corresponding
interaction Hamiltonians are discussed.  In Sec.~\ref{sec:3} we
briefly summarise the realization of a coherent micro-maser-type
dynamics, and in Sec.~\ref{sec:4} the incoherent regime is introduced.
A trapping mechanism that works also in the incoherent regime is then
discussed in Sec.~\ref{sec:5}. Finally, a summary and some conclusions
are given in Sec.~\ref{sec:6}.

\section{Laser excitation scheme for the trapped ion} \label{sec:2}
The centre-of-mass coordinate of a single ion bound in a
radio-frequency trap, such as for example a Paul trap, is subject to a
dynamics that can be described to good approximation as a 3D harmonic
oscillation with three characteristic vibrational frequencies along
the principal axes of the trapping potential. The geometry of the
propagation of externally applied laser fields can be configured in
such a way as to influence only the ion's centre-of-mass motion in the
direction of one principal axis, that is associated with the trap
frequency $\nu$. Thus, one may consider the system as being
one-dimensional, due to the natural decoupling from the remaining two
other degrees of freedom~\cite{eschner}.

In what follows, we always consider the case of quasi-monochromatic,
collimated laser beams, whose wave-vectors are parallel to the chosen
principal axis of the trap, that we may specify by the centre-of-mass
coordinate $x$. In the case where laser-atom interactions are
implemented by two-photon induced Raman transitions, the wave-vector
of the beat note of the required two laser beams shall point along
this axis.

\begin{figure}
  \begin{center}
    \epsfig{file=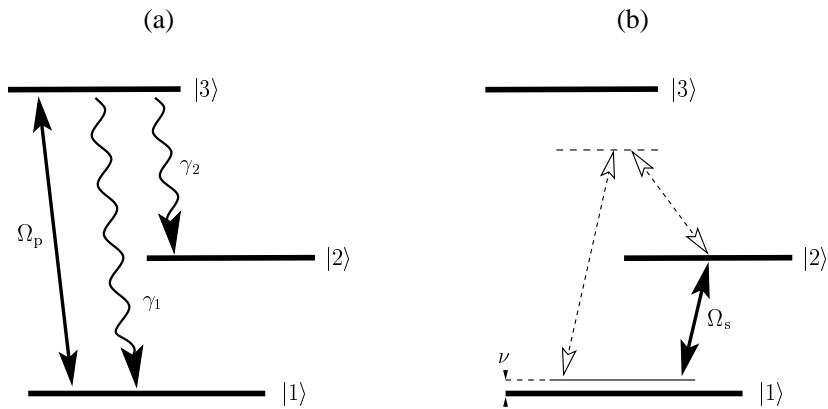,scale=0.75}
    \caption{Laser-excitation scheme for the preparation of trapping
      state. During the optical-pumping interaction (a) a laser (Rabi
      frequency $\Omega_{\rm p}$) together with spontaneous emissions
      (rates $\gamma_{1,2}$) prepares the system in state $|2\rangle$.
      In the sideband interaction (b) the first vibrational sideband
      (vibration frequency $\nu$) is driven (Rabi frequency
      $\Omega_{\rm s}$) either directly (solid arrow) or in a Raman
      configuration (dashed arrows).}
    \label{fig:scheme}
  \end{center}
\end{figure}

The laser-excitation scheme shall consist of cycles, each being
implemented by two interactions, an optical-pumping interaction, and a
sideband interaction. The optical pumping is intended to invert the
populations of the two states $|1\rangle$ and $|2\rangle$, cf.
Fig.~\ref{fig:scheme}a.
It is implemented via the laser-driven electric-dipole transition
$|1\rangle \!\leftrightarrow\! |3\rangle$, and the spontaneous decay
on the electric-dipole transition $|3\rangle \!\rightarrow\!
|2\rangle$.  The equation of motion, governing the dynamics of the
vibronic density operator of the ion, $\hat{\varrho}$, is then given
by
\begin{eqnarray}
  \label{eq:master-pump}
  \dot{\hat{\varrho}} & = & - \frac{i}{\hbar} [ \hat{H}_0 \!+\!
  \hat{H}_{\rm p}(t), \hat{\varrho} ] 
  + \gamma_1 \left( \hat{A}_{13} \, \hat{\mathcal R}_1 [\hat{\varrho}] 
    \, \hat{A}_{31} 
    - {\textstyle\frac{1}{2}} \hat{A}_{33} \, \hat{\varrho} 
    - {\textstyle\frac{1}{2}} \hat{\varrho} \, \hat{A}_{33} \right)
  \nonumber \\ 
  & & + \, \gamma_2 \left( \hat{A}_{23} \, \hat{\mathcal R}_2 
    [\hat{\varrho}] \, \hat{A}_{32}
  - {\textstyle\frac{1}{2}} \hat{A}_{33} \, \hat{\varrho} -
  {\textstyle\frac{1}{2}} \hat{\varrho} \,
  \hat{A}_{33} \right) .
\end{eqnarray}
The operators $\hat{A}_{ij} \!=\! |i\rangle \langle j|$ induce
transitions between the ion's electronic states $|j \rangle$ and
$|i\rangle$ and $\gamma_1$ and $\gamma_2$ are the spontaneous decay
rates of state $|3\rangle$ decaying to states $|1\rangle$ and
$|2\rangle$, respectively. The free Hamiltonian $\hat{H}_0$ reads
\begin{equation}
  \label{eq:H0}
  \hat{H}_0 = \hbar\nu \, \hat{a}^\dagger \hat{a} + \sum_{i=1}^3
  \hbar\omega_i \hat{A}_{ii} ,
\end{equation}
where $\hat{a}$ and $\hat{a}^\dagger$ are the annihilation and
creation operators of vibrational quanta, respectively.

The laser interaction driving the optical pump process reads
\begin{equation}
  \label{eq:pump-part}
  \hat{H}_{\rm p}(t) = {\textstyle\frac{1}{2}} \, \hbar \Omega_{\rm p}
  \, \hat{A}_{31} \, e^{-i \omega_{\rm p} t} + {\rm H.a.} ,
\end{equation}
where $\Omega_{\rm p}$ is the Rabi frequency denoting the coupling
strength of the pump laser beam with the electric-dipole moment of the
ion. The pump-laser frequency $\omega_{\rm p}$ shall be near resonant
with the electronic transition frequency $\omega_{31}$ of the states
$|1\rangle$ and $|3\rangle$. For this interaction the laser beam is
supposed to propagate along a direction perpendicular to the motional
direction $x$ under consideration, to avoid coupling of the electronic
and vibrational degrees of freedom in $x$ direction.

The photonic recoil acting on the centre-of-mass motion of the ion
during the spontaneous emissions is given by the operator $\hat{\mathcal
  R}_i$ ($i\!=\! 1,2$)
\begin{equation}
  \label{eq:recoil-op1}
  \hat{\mathcal{R}}_i [\hat{\varrho}] = \int_{-1}^1 ds \,
    w(s) \, e^{i s \eta_i (
    \hat{a}^\dagger + \hat{a} )} \, \hat{\varrho} \,
  e^{-i s \eta_i ( \hat{a}^\dagger + \hat{a} )} ,
\end{equation}
with $w(s) \!=\!  \frac{3}{8}(1\!+\!s^2)$ describing the dipole
radiation characteristics and the Lamb--Dicke parameters $\eta_1$ and
$\eta_2$ correspond to the spontaneous transitions $|3\rangle \!\to\!
|1\rangle$ and $|3\rangle \!\to\! |2\rangle$, respectively.  For
simplicity we neglect, for the moment, the vibrational scattering
during the spontaneous photon emissions, i.e. we choose $\hat{\mathcal
  R}_i [\hat{\varrho}] \!\to\! \hat{\varrho}$, though later these
effects will be taken into account to full extent numerically.

The stationary solution of the optical pumping process is
approximately reached after a sufficiently long time of laser
interaction. In terms of electronic-state matrix elements
$\hat{\rho}_{ij} \!=\! \langle i | \hat{\varrho} |j \rangle$ of the
density operator, that are still operators acting on the vibrational
degree of freedom, the stationary solution reads
\begin{eqnarray}
  \label{eq:stat-pump}
  \hat{\rho}_{12}(\tau_{\rm p}) & = &  \hat{\rho}_{23}(\tau_{\rm p}) 
  = \hat{\rho}_{31}(\tau_{\rm p}) \approx 0 \\
  \hat{\rho}_{11}(\tau_{\rm p}) & = & \hat{\rho}_{33}(\tau_{\rm p})
  \approx 0 \\
  \label{eq:stat-pump22}
  \hat{\rho}_{22}(\tau_{\rm p}) & = & \hat{U}_0(\tau_{\rm p}) \left[
  \hat{\rho}_{11}(0) + \hat{\rho}_{22}(0) \right]
  \hat{U}_0^\dagger(\tau_{\rm p}) .
\end{eqnarray}
Here we assume that initially, due to relaxation, the population in
state $|3\rangle$ is zero, i.e., $\hat{\rho}_{33}(0) \!=\! 0$.  Thus,
all population from state $|1\rangle$ merges with the population of
state $|2\rangle$, while the ion is subject to oscillation in the trap
potential.

The subsequent sideband interaction is implemented either by direct or
by Raman excitation, resonant to the red vibrational sideband of the
weak transition $|1\rangle \!\leftrightarrow\!  |2\rangle$, cf.
Fig.~\ref{fig:scheme}b. Its interaction Hamiltonian reads
\begin{equation}
  \label{eq:Hs-1}
  \hat{H}_{\rm s}(t) = {\textstyle\frac{1}{2}} \hbar \Omega \,
    \hat{A}_{21} \, e^{i ( k \hat{x} - \omega t)} + {\rm H.a.} ,
\end{equation}
where $k$ and $\omega$ are the projection of the wave-vector on the
$x$ axis and the frequency, respectively, either of the laser field
for direct excitation, or of the beat note of the Raman lasers.
Moreover, the Rabi frequency $\Omega$ denotes the coupling strength of
this transition.  For the case of resonance with the red vibrational
sideband, $\omega \!=\! \omega_{21} \!-\! \nu$, this laser interaction
induces resonant vibronic transitions between states $|2\rangle
|n\rangle$ and $|1\rangle |n\!+\!1\rangle$, where $|n\rangle$ are
energy eigenstates of the centre-of-mass vibration in the trap
potential. For this case the Hamiltonian~(\ref{eq:Hs-1}) simplifies,
in rotating-wave approximation and in the interaction picture, to a
nonlinear Jaynes--Cummings type Hamiltonian~\cite{njcm},
\begin{equation}
  \label{eq:njc-ham2}
  \hat{H}_{\rm s} = \frac{1}{2} \hbar \Omega_{\rm s} \hat{A}_{21}
  \hat{f}(\hat{n}; \eta) \, \hat{a} + {\rm H.a.} \; ,
\end{equation}
where $\Omega_{\rm s} \!=\! i\eta\Omega$ is the Rabi frequency on the
red vibrational sideband in the Lamb--Dicke limit.  The effects that
arise in the regime beyond the Lamb--Dicke approximation, i.e., with
rather large Lamb--Dicke parameter $\eta$ are described by the
excitation-dependent operator function $\hat{f}(\hat{n}; \eta)$ with
$\hat{n}$ being the number operator of vibrational quanta in $x$
direction.  Note, that for a Raman configuration, changing the
laser-beam propagation geometry varies the projection $k$ of the
beat-note wave-vector on the $x$ axis. Thus, the Lamb--Dicke parameter
can be tuned up to rather large values.

The operator function $\hat{f}(\hat{n}; \eta)$ depends solely on the
number of vibrational quanta and can be given by a normally ordered
expression as
\begin{eqnarray}
  \label{eq:f-func}
  \hat{f}(\hat{n}; \eta) & = & e^{-\eta^2/2}
  \sum_{n=0}^\infty (-1)^n \frac{\eta^{2n}}{n! \, (n+1)!} \,
  \hat{a}^{\dagger n} \hat{a}^n \\ \nonumber & = & : \, \left( \eta
    \sqrt{\hat{a}^\dagger \hat{a}} \right)^{-1} J_1 \left( 2 \eta
    \sqrt{\hat{a}^\dagger \hat{a}} \right) \, : \, e^{-\eta^2/2} ,
\end{eqnarray}
where $:\,\,:$ denotes normal ordering, and $J_k(z)$ are the Bessel
functions of integer order. Clearly these functions introduce a
dependence of the laser-ion coupling on the vibrational excitation of
the ion. They can be understood as analogy to Franck--Condon overlap
integrals, however, in momentum space due to the recoil of absorbed
and emitted photons.

For this approximation, i.e. neglecting spontaneous recoil effects, in
each cycle of optical pumping and sideband interaction, the mean
vibrational excitation will be increased or at least held constant.
The dominant processes in one cycle can thus be given by the cascade
\begin{displaymath}
  \begin{array}{l} 
    |1\rangle |n\rangle \\
    |2\rangle |n\rangle 
  \end{array}
  \stackrel{\rm pump}{\Longrightarrow} 
  |2\rangle |n\rangle 
  \stackrel{\rm sideband}{\Longrightarrow}
  \begin{array}{l}
    |2\rangle |n\rangle \\ 
    |1\rangle |n\!+\!1\rangle 
  \end{array} ,
\end{displaymath}
thus vibrational transitions $|n\rangle \!\rightarrow\!  |n\rangle$
and $|n\rangle \!\rightarrow\! |n\!+\!1\rangle$ are performed in each
cycle, leading to a net increase of the vibrational excitation.

\section{Coherent time evolution} \label{sec:3}
In the absence of decoherence effects the time evolution for the
sideband interaction is described by a unitary time-evolution operator
$\hat{U}(t)$.  Thus, starting with the vibronic density operator
$\hat{\varrho}$ of the trapped ion [cf. Eq.~(\ref{eq:stat-pump22})]
\begin{equation}
  \label{eq:pumped2}
  \hat{\varrho}(t\!+\! \tau_p) = \hat{A}_{22} \, \hat{U}_0(\tau_p) \,
  \left[ \hat{\varrho}_{11}(t) + \hat{\varrho}_{22}(t) \right] \,
  \hat{U}_0^\dagger(\tau_p) ,
\end{equation}
at the time $t \!+\! \tau_p$, right after the optical pumping, the
density operator reads after an interaction time $\tau_s$ of the
sideband interaction,
\begin{equation}
  \label{eq:step-op}
  \hat{\varrho}(t \!+\! \tau_p \!+\! \tau_s) = \hat{U}_{\rm s}(\tau_s) \,
  \hat{\varrho}(t \!+\! \tau_p) \, \hat{U}_{\rm s}^\dagger(\tau_s) ,
\end{equation}
where $\hat{U}_{\rm s}(t) \!=\! \hat{U}_{0}(t) \exp( -i \hat{H}_{\rm
  s} t/\hbar) $ and $\hat{U}_{0}(t) \!=\! \exp( -i \hat{H}_0 t/\hbar)
$.

Starting from an initial density operator at time $t_k$ we obtain the
density operator at the time $t_{k+1} \!=\! t_k \!+\! \tau_p \!+\!
\tau_s$, after a complete cycle of pump and sideband interaction, by
insertion of Eq.~(\ref{eq:pumped2}) into (\ref{eq:step-op}) as
\begin{equation}
  \label{eq:rho-complete}
  \hat{\varrho}(t_{k+1}) = \hat{U}_{\rm s}(\tau_s) \, |2 \rangle \,
  \hat{U}_0(\tau_p) \left[ \langle 1 | \hat{\varrho}(t_k) | 1
  \rangle \!+\! \langle 2 | \hat{\varrho}(t_k) | 2
  \rangle \right]
  \hat{U}_0^\dagger(\tau_p) \, \langle 2 | \, \hat{U}_{\rm s}^\dagger(\tau_s)
\end{equation}
The vibrational number statistics is obtained from the density
operator as
\begin{equation}
  \label{eq:pn}
  P_n(t) = {\rm Tr} \left[ \hat{\varrho}(t) \, |n\rangle\langle n|
  \right] ,
\end{equation}
Using Eqs~(\ref{eq:rho-complete}) and (\ref{eq:pn}) straightforward
calculation~\cite{review} results in a recurrence relation for the
probabilities of vibrational quantum numbers
\begin{equation}
  \label{eq:recurrence}
  P_n(t_{k+1}) = w_{\rm coh}(n|n \!-\! 1) \, P_{n-1}(t_k) +
  w_{\rm coh}(n|n) \, P_n(t_k) ,
\end{equation}
where the probabilities for excitation or survival are characterised
by a coherent sideband-interaction and read
\begin{eqnarray}
  \label{eq:coh-rates}
  w_{\rm coh}(n \!+\! 1|n) & = & \sin^2 \! \left[ {\textstyle\frac{1}{2}} 
    |\Omega_{\rm s}|\tau_{\rm s} \, f(n; \eta) \sqrt{n \!+\! 1}
  \right] , \\
  \label{eq:coh-rates-exc}
  w_{\rm coh}(n|n) & = & \cos^2 \! \left[ {\textstyle\frac{1}{2}} 
    |\Omega_{\rm s}| \tau_{\rm s} \, f(n; \eta) \sqrt{n \!+\! 1} \right] ,
\end{eqnarray}
where we have used the function $f(n;\eta) \!=\! \langle n |
\hat{f}(\hat{n};\eta) |n\rangle$, which can be derived from
Eq.~(\ref{eq:f-func}) as Laguerre polynomial
\begin{equation}
  \label{eq:fn}
  f(n;\eta) = (n \!+\! 1)^{-1} L_n^{(1)}(\eta^2) \, e^{-\eta^2/2} .
\end{equation}

Equation~(\ref{eq:recurrence}) describes the change of the vibrational
statistics depending on the number of cycles performed. This
recurrence relation contains the familiar trapping mechanism known
from micro-maser theory~\cite{filipowicz}: The trapping state
$|n_0\rangle$ is reached when the excitation rate (\ref{eq:coh-rates})
from state $|n_0\rangle$ to $|n_0\!+\!1\rangle$ vanishes, so that all
the population accumulates in state $|n_0\rangle$.  The condition
$w_{\rm coh}(n_0 \!+\! 1|n_0) \!=\!  0$ thus leads to the condition
for $|n_0\rangle$ being a trapping state:
\begin{equation}
  \label{eq:trap-cond}
  |\Omega_{\rm s}| \tau_{\rm s} \, f(n_0; \eta) \, \sqrt{n_0 \!+\! 1} = 
  2 \pi m , \quad m \in {\Bbb Z} .
\end{equation}

In the Lamb--Dicke limit the nonlinear coupling function in
Eq.~(\ref{eq:trap-cond}) reduces to $\lim_{\eta\rightarrow 0} f(n;
\eta) \!=\! 1$, and Eq.~(\ref{eq:trap-cond}) reduces to the familiar
trapping-state condition for the micro-maser case~\cite{vaser}. This
case is characterised by a non-vanishing function $f(n_0;\eta)$, upon
which the condition for the interaction time results
\begin{equation}
  \label{eq:standard-cond}
  \tau_{\rm s} = \frac{2\pi m}{|\Omega_{\rm s}| 
    f(n_0; \eta) \sqrt{n_0 \!+\! 1}} 
  , \quad m \in {\Bbb Z} .
\end{equation}
This type of trapping states is generated by a coherent mechanism,
that consists of complete Rabi cycles at the trapping-state condition,
leading to a locked vibrational quantum number.

Eq.~(\ref{eq:trap-cond}) on the other hand, contains also a quite
distinct type of trapping condition, namely that one where the
Franck--Condon-type nonlinearity in the laser-ion coupling strength
leads to a vanishing coupling, i.e., if
\begin{equation}
  \label{eq:cond-nl}
  f(n_0; \eta) = 0 .
\end{equation}
This case does not correspond to complete Rabi cycles, but represents
the breakdown of the laser-ion coupling mechanism, due to
non-overlapping trap eigenstates in momentum representation, that are
shifted off each other by the differing amount of the effective
photonic momentum $\hbar k$.  This novel type of trapping states may
be set up in both a coherent or incoherent way.

Let us illustrate the qualitative differences between the two types of
trapping states for a coherent sideband interaction and consider the
two different ways of generating a trapping state $|n_0\rangle$ with
$n_0 \!=\! 50$ vibrational quanta. The well known way is to use the
standard trapping-state condition~(\ref{eq:standard-cond}) to fix the
pulse area $|\Omega_{\rm s}| \tau_s$ of the laser resonant with the
vibrational sideband for a given Lamb--Dicke parameter $\eta$. 
The alternative way of generating this trapping state in a coherent
fashion is given by adjusting the laser-beam geometry for sideband
interaction (in a Raman configuration) in such a way as to tune to a
Lamb--Dicke parameter that fulfils condition~(\ref{eq:cond-nl}). The
laser-pulse area is then chosen in such a way, that lower-lying
trapping states due to complete Rabi cycles are avoided. 

\begin{figure}
  \begin{center}
    \epsfig{file=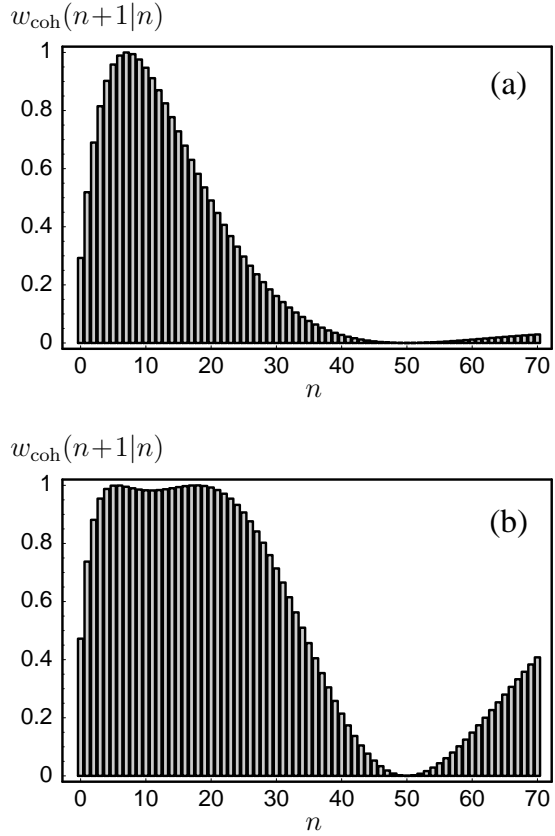,scale=0.9}
    \caption{Coherent transition rates $w_{\rm coh}(n\!+\! 1|n)$  
      for (a) $\eta \!=\! 0.1$, $|\Omega_{\rm s}| \tau_{\rm s} \!=\!
      1.149$ [trapping-state condition (\ref{eq:standard-cond}) for
      $n_0 \!=\! 50$], and for (b) $\eta \!=\! 0.268$ [trapping-state
      condition (\ref{eq:cond-nl}) for $n_0 \!=\!  50$] and
      $|\Omega_{\rm s}| \tau_{\rm s} \!=\! \pi/2$.}
    \label{fig:coh-rate}
  \end{center}
\end{figure}

The improvement is illustrated when comparing the coherent transition
rates $w_{\rm coh}(n \!+\! 1|n)$ in the two cases,
cf.~Fig.~\ref{fig:coh-rate}.
The dip of vanishing transition rate around the trapping state $|n_0
\!=\! 50\rangle$ is sharper for the rate shown in part (b), which is
due to the nonlinear modification encoded in the function $f(n;\eta)$
when $\eta$ is not too small.  As a consequence, when the quantum
state approaches the trapping state $|n_0\!=\! 50\rangle$, the
dynamics is significantly slowed down in case (a) since the rates
$w_{\rm coh}(n\!+\! 1|n)$ are smaller (for $n \!=\! n_0 \!-\! 1, n_0
\!-\! 2, n_0 \!-\! 3$) as in case (b). In both cases the transition
rates may be suppressed at certain vibrational numbers below the
trapping state. These coherent effects emerge when the ion performs
nearly complete Rabi cycles on the electronic transition at certain
values of the integer $m$ in the trapping-state
condition~(\ref{eq:standard-cond}).  Such subharmonic resonances
degrade the generation of the trapping state, since they substantially
decrease the flow of population into the trapping state.

The sensitivity to subharmonic resonances seems a handicap to both
coherent ways of achieving a trapped state, in addition to the
fundamental problem of preserving coherence. The novel scheme,
however, evades both problems, since it is applicable in an incoherent
way.

\section{Incoherent dynamics} \label{sec:4}
The trapping mechanisms described above, being based on coherent
temporal evolution, are disturbed by any decoherence, as for example
the phase fluctuations of the Raman lasers, or spontaneous emissions.
It would be favourable to have a trapping mechanism at hand, that does
not depend on the coherence of the underlying physical processes and
is therefore not disturbed by such effects.  Then, a trapping state
might be reached even under the influence of various decohering
perturbations and would be more robust than the micro-maser-type
trapping states.

We consider the sideband interaction under the influence of
decoherence, i.e. electronic dephasing, such as produced by
phase-fluctuating lasers, spontaneous emissions, or collisions with
background vapour.  The total amount of phase fluctuations will be
described by a dephasing rate $\gamma$ that gradually destroys the
coherence between electronic levels $|1\rangle$ and $|2\rangle$.  For
simplicity we do not consider electronic dephasing mechanisms that
depend on the vibrational excitation. Such effects may appear due to
laser fluctuations~\cite{schneider}, spontaneous Raman
processes~\cite{difidio1,difidio2}, or combinations of
both~\cite{budini}. Moreover, for the analytic treatment we neglect
the spontaneous recoil effects. These however will be included in the
complete numerical solution shown later.

The master equation for the sideband-coupling interaction, including
the decoherent phase fluctuations reads in the interaction picture as
\begin{equation}
  \label{eq:master}
  \dot{\hat{\varrho}} = - \frac{i}{\hbar} [ \hat{H}_{\rm s} ,
  \hat{\varrho} ] 
  - \frac{\gamma}{2} \left( \hat{A}_{11} \, \hat{\varrho} \,
    \hat{A}_{22} + \hat{A}_{22} \, \hat{\varrho} \,
    \hat{A}_{11} \right) ,
\end{equation}
where the second term accounts for the electronic
dephasing~\cite{dephasing1,dephasing2}. The equations of motion for
the electronic populations are derived from Eq.~(\ref{eq:master}) as
\begin{eqnarray}
  \label{eq:master-11}
  \dot{\hat{\varrho}}_{11} & = & -
  \frac{i}{2} \left( \Omega_{\rm s}^\ast \, \hat{b}^\dagger \, 
    \hat{\varrho}_{21} - \Omega_{\rm s} \,
    \hat{\varrho}_{12} \, \hat{b} \right) , \\
  \label{eq:master-22}
  \dot{\hat{\varrho}}_{22} & = & -
  \frac{i}{2} \left( \Omega_{\rm s} \, \hat{b} \,
    \hat{\varrho}_{12} - \Omega_{\rm s}^\ast \,
    \hat{\varrho}_{21} \, \hat{b}^\dagger \right) ,
\end{eqnarray}
where we have made use of the nonlinearly deformed annihilation
operator
\begin{equation}
  \label{eq:b-op-def}
  \hat{b} = \hat{f}(\hat{a}^\dagger \hat{a}; \eta) \, \hat{a} .
\end{equation}
The time evolution of the electronic coherences, on the other hand, is
modified
\begin{eqnarray}
  \label{eq:master-12}
  \dot{\hat{\varrho}}_{12} & = & -
  \frac{\gamma}{2} \, \hat{\varrho}_{12} - \frac{i}{2}
  \Omega_{\rm s}^\ast \left( \hat{b}^\dagger \hat{\varrho}_{22} -
    \hat{\varrho}_{11} \, \hat{b}^\dagger \right) , \\
  \label{eq:master-21}
  \dot{\hat{\varrho}}_{21} & = & -
  \frac{\gamma}{2} \, \hat{\varrho}_{21} - \frac{i}{2}
  \Omega_{\rm s} \left( \hat{b} \, \hat{\varrho}_{11} -
    \hat{\varrho}_{22} \, \hat{b} \right) .
\end{eqnarray}

Let us now consider the incoherent regime where the dephasing rate
$\gamma$ is larger than the Rabi frequency on the vibrational
sideband: $\gamma \!>\! |\Omega_{\rm s}|$.  Nevertheless, we consider
the case of well-resolved sidebands, $\nu \!>\! \gamma$, and thus
$\gamma$ is in the range given by
\begin{equation}
  \label{eq:inc-range}
  |\Omega_{\rm s}| < \gamma < \nu . 
\end{equation}
Then the rotating-wave approximation with respect to $\nu$ for the
derivation of Eq.~(\ref{eq:njc-ham2}) is valid.  For typical
vibrational frequencies of $\nu/2\pi \!\approx\! 1 \!\ldots\! 10$ MHz
and Rabi frequencies of $|\Omega|/2\pi \!\approx\! 10 \!\ldots\!  100$
kHz, the dephasing rate is supposed on the order of $\gamma/2\pi
\!\approx\! 100\mbox{ kHz} \ldots\! 1\mbox{ MHz}$.  In this incoherent
regime adiabatic elimination of the electronic coherences can be
performed by neglecting the time derivatives in
Eqs.~(\ref{eq:master-12}) and (\ref{eq:master-21}) and solving for the
electronic coherences.  Inserting the resulting electronic coherences
into the equations of motion for the electronic populations
(\ref{eq:master-11}) and (\ref{eq:master-22}) we obtain
\begin{eqnarray}
  \label{eq:adiab-11}
  \dot{\hat{\varrho}}_{11} & = &
  \frac{|\Omega_{\rm s}|^2}{\gamma} \left( \hat{b}^\dagger
    \hat{\varrho}_{22} \, \hat{b} - {\textstyle\frac{1}{2}} 
    \hat{b}^\dagger \hat{b}
    \, \hat{\varrho}_{11} - {\textstyle\frac{1}{2}} \hat{\varrho}_{11}
    \, \hat{b}^\dagger \hat{b} \right) , \\
  \label{eq:adiab-22}
  \dot{\hat{\varrho}}_{22} & = &
  \frac{|\Omega_{\rm s}|^2}{\gamma} \left( \hat{b} \,
    \hat{\varrho}_{11} \, \hat{b}^\dagger - {\textstyle\frac{1}{2}} \hat{b}
    \hat{b}^\dagger \hat{\varrho}_{22} - {\textstyle\frac{1}{2}}
    \hat{\varrho}_{22} \, \hat{b} \hat{b}^\dagger \right) .
\end{eqnarray}

These equations can be decoupled and solved by use of the
dressed-state coefficients
\begin{equation}
  \label{eq:c-def}
  C_{mn}^{(\pm)} = \langle m | \hat{\varrho}_{11} | n
  \rangle \pm \langle m \!-\! 1 | \hat{\varrho}_{22} | n
  \!-\! 1 \rangle .
\end{equation}
From Eqs.~(\ref{eq:adiab-11}) and (\ref{eq:adiab-22}) we obtain the
decoupled differential equations for the coefficients
$C_{mn}^{(\pm)}(t)$ as
\begin{equation}
  \label{eq:c-diff}
  \dot{C}_{mn}^{(\pm)} = - {\textstyle\frac{1}{4}}
  \left[ \sqrt{\gamma_{m\!-\!1}(\eta)} \mp
  \sqrt{\gamma_{n\!-\!1}(\eta)} \right]^2 C_{mn}^{(\pm)} ,
\end{equation}
where the excitation-dependent damping coefficients are given by
\begin{equation}
  \label{eq:gamma}
  \gamma_n(\eta) = 2 \gamma s \, (n \!+\! 1) \,
  [ f(n; \eta) ]^2 ,
\end{equation}
and the saturation parameter reads $s \!=\! 2 |\Omega_{\rm s}|^2 /
\gamma^2$. Using Eq.~(\ref{eq:fn}) the nonlinear damping
coefficients can be given in terms of Laguerre polynomials as
\begin{equation}
  \label{eq:gamma-laguerre}
  \gamma_n(\eta) = \frac{2 \gamma s}{n\!+\!1} \,
  |L_n^{(1)}(\eta^2)|^2 \, e^{-\eta^2} .
\end{equation}
Note, that the laser-ion coupling nonlinearity appears only in these
damping coefficients, and that in the Lamb--Dicke approximation they
become linear with $n$:
\begin{equation}
  \label{eq:limes}
  \lim_{\eta\rightarrow 0} \gamma_n(\eta) = 2\gamma s \, (n \!+\! 1) .
\end{equation}

The vibrational number statistics $P_n$ can be easily obtained from
the matrix elements $C_{mn}^{(\pm)}$ as follows:
\begin{equation}
  \label{eq:reduced-densmat}
  P_n = {\textstyle\frac{1}{2}} \left( C_{nn}^{(+)} \!+\!  C_{nn}^{(-)}
    \!+\!  C_{n+1,n+1}^{(+)} \!-\!  C_{n+1,n+1}^{(-)} \right) .
\end{equation}
Thus only the diagonal matrix elements $C_{nn}^{(\pm)}$ are needed,
which obey the equations of motion
\begin{equation}
  \label{eq:diag-diff}
  \dot{C}_{nn}^{(+)} = 0 , \quad
  \dot{C}_{nn}^{(-)} = - \gamma_{n\!-\!1}(\eta) \, C_{nn}^{(-)} .
\end{equation}
Since the incoherent dynamics of the sideband interaction follows the
optical pumping, the initial conditions for the density-matrix
elements can be written as
\begin{eqnarray}
  \label{eq:initial}
  \langle n | \hat{\varrho}_{11}(t_k \!+\! \tau_{\rm p}) | n \rangle &
  = & 0 , \\ \langle n | \hat{\varrho}_{22}(t_k \!+\! \tau_{\rm p}) | n
  \rangle & = & P_n(t_k \!+\! \tau_{\rm p}) = P_n(t_k) ,
\end{eqnarray}
where again $t_k$ denotes the time after a complete cycle, and $\tau_p$
is the pump time. Using these initial conditions one obtains, via
Eq.~(\ref{eq:c-def}), the initial conditions for the coefficients and
the solution of Eq.~(\ref{eq:diag-diff}) results as
\begin{eqnarray}
  \label{eq:sol+}
  C_{nn}^{(+)}(t_{k+1}) & = & P_{n-1}(t_k) , \\
  \label{eq:sol-}
  C_{nn}^{(-)}(t_{k+1}) & = & - P_{n-1}(t_k) \, \exp [ -
    \gamma_{n\!-\!1}(\eta) \, \tau_{\rm s} ] ,
\end{eqnarray}
Inserting Eqs.~(\ref{eq:sol+}) and (\ref{eq:sol-}) into
Eq.~(\ref{eq:reduced-densmat}) one finally obtains a modified
recurrence relation for the number statistics after subsequent cycles
in the incoherent regime
\begin{equation}
  \label{eq:recurrence-inc}
  P_{n}(t_{k+1}) = w_{\rm inc}(n|n\!-\!1) \,
  P_{n\!-\!1}(t_k) + w_{\rm inc}(n|n) \, P_n(t_k)  ,
\end{equation}
with the transition rates in the incoherent regime of sideband
interaction, 
\begin{eqnarray}
  \label{eq:cond-prob1}
  w_{\rm inc}(n\!+\!1|n) & = & {\textstyle\frac{1}{2}} \left\{ 1 - \exp[ -
      \gamma_n(\eta) \, \tau_{\rm s} ] \right\} , \\
  \label{eq:cond-prob2}
  w_{\rm inc}(n|n) & = & {\textstyle\frac{1}{2}} \left\{ 1 + \exp[ -
      \gamma_n(\eta) \, 
      \tau_{\rm s} ] \right\} .
\end{eqnarray}

\section{Incoherent trapping mechanism} \label{sec:5}
Let us now study in more detail the rates $w_{\rm inc}(n \!+\!1 |n)$,
that determine the increase of the mean vibrational excitation. If
those rates are non-vanishing, the populations of vibrational quantum
numbers will be shifted to the next higher lying quantum numbers in
each cycle, leading to an overall increase of the mean vibrational
excitation. If, however, $w_{\rm inc}(n_0\!+\!1|n_0)$ is zero for some
vibrational number $n_0$, the recurrence
relation~(\ref{eq:recurrence-inc}) has a cutoff at that number $n_0$.
That is, all populations that were initially at vibrational quantum
numbers below $n_0$ will be accumulated after some time in the
vibrational state $|n_0\rangle$. Clearly, populations in levels higher
than $n_0$ will be shifted to still higher quantum numbers in each
cycle.

Thus two distinct dynamical regimes can be found in the incoherent
dynamics: A regime where the mean vibrational excitation is increased
in each cycle, and a trapping regime for populations initially below a
certain vibrational quantum number $n_0$ where the condition $w_{\rm
  inc}(n_0 \!+\! 1|n_0) \!=\! 0$ holds. The latter we will call, in
the following, {\it incoherent trapping states} since they emerge from
a dynamics bare of any coherence.

In fact, incoherent trapping states emerge from a breakdown of the
ion-laser coupling and are thus related to the second type of trapping
states with coherent dynamics discussed before. To see this relation,
let us consider the condition for an incoherent trapping state at
vibrational quantum number $n_0$:
\begin{equation}
  \label{eq:zeros}
  w_{\rm inc}(n_0\!+\!1|n_0) = 0 = 
  {\textstyle\frac{1}{2}} \left\{ 1 - \exp \left[ -
      \gamma_{n_0}(\eta) \, \tau_{\rm s} \right] \right\} ,
\end{equation}
which is equivalent to the condition 
\begin{equation}
  \label{eq:cond-inc}
  \gamma_{n_0}(\eta) = 0 .
\end{equation}
From their definition~(\ref{eq:gamma}) follows $\gamma_{n_0}(\eta)
\!\propto\! [f(n_0; \eta)]^2$, such that the condition for incoherent
trapping states~(\ref{eq:cond-inc}) is equivalent to the condition of
nonlinear coherent trapping states~(\ref{eq:cond-nl}). This is because
both types of trapping states rely on the same mechanism, namely, the
breakdown of the laser-ion coupling strength due to the vanishing of
Franck--Condon type overlap integrals of vibrational wave functions.
Thus, we deal with a trapping mechanism that appears both in the
coherent and the incoherent regime and therefore does not rely on
coherent dynamics.

Incoherent trapping states may be conveniently prepared with no need
for coherence. Decoherence effects will not prevent the emergence of
these trapping states. For small vibrational quantum numbers one
typically would need only few cycles such that little decoherence is
accumulated.  However, for generating highly excited vibrational
number states, the mechanism of incoherent trapping states is
superior. Then a rather large number of cycles is required, where in
each cycle decoherence effects would eventually prevent any coherent
mechanism from working.

Let us now study the dynamic increase of the vibrational excitation
in the incoherent regime.  For large saturation $s$ and/or large
interaction times $\tau_{\rm s}$, the transition rates $w_{\rm inc}$
can be approximated as
\begin{eqnarray}
  \label{eq:cond-prob-approx1}
  w_{\rm inc}(n\!+\!1|n) & \approx & \frac{1}{2} 
  \Big[ 1 - \sum_{n_0(\eta)} \delta_{n,n_0(\eta)} \Big] , \\
  \label{eq:cond-prob-approx2}
  w_{\rm inc}(n|n) & \approx & \frac{1}{2} 
  \Big[ 1 + \sum_{n_0(\eta)} \delta_{n,n_0(\eta)} \Big] ,
\end{eqnarray}
where $\delta_{n,m}$ denotes the Kronecker delta symbol and the sum
extends over all possible trapping numbers $n_0(\eta)$ at the given
Lamb--Dicke parameter $\eta$.  This form of the transition rates is an
excellent approximation for a wide range of parameters and holds for
the following examples.  Thus, in regions of vibrational numbers $n
\!\neq\! n_0(\eta)$, the transition rates simplify to
\begin{equation}
  \label{eq:approx-inc}
  w_{\rm inc}(n \!+\! 1|n) \approx w_{\rm inc}(n|n) \approx \frac{1}{2} ,
\end{equation}
and the recurrence relation~(\ref{eq:recurrence-inc}) for the
incoherent dynamics reduces to
\begin{equation}
  \label{eq:binom-recurrence}
  P_n(t_{k+1}) = {\textstyle\frac{1}{2}} \, P_n(t_k) 
  + {\textstyle\frac{1}{2}} \, P_{n-1}(t_k) .
\end{equation}
Due to the constant transition rates, now the evolution towards a
trapping state is not slowed down by quasi-trapping conditions as may
be found in the coherent rates.

Using Eq.~(\ref{eq:binom-recurrence}) it can be shown that, far from
any trapping state, the mean vibrational excitation and variance obey
the following dynamics:
\begin{eqnarray}
  \label{eq:mean}
  \langle \hat{n}(t_{k+1}) \rangle & = & \langle \hat{n}(t_k) \rangle
  + {\textstyle\frac{1}{2}} , \\
  \label{eq:variance} 
  \langle \left[ \Delta \hat{n}(t_{k+1}) \right]^2 \rangle & = &
  \langle \left[ \Delta \hat{n}(t_k) \right]^2 \rangle +
  {\textstyle\frac{1}{4}}  .
\end{eqnarray}
These relations show that in each cycle, on the average, half a
vibrational quantum is created in the ionic motion.  Moreover, the
mean excitation increases faster than the variance so that for a large
number of cycles $k$ the relative variance $\langle \Delta\hat{n}^2
\rangle / \langle \hat{n} \rangle$ reaches the stationary value,
\begin{equation}
  \label{eq:stat-var}
  \lim_{k\rightarrow\infty} \frac{\langle \left[ \Delta \hat{n}(t_k)
    \right]^2 \rangle}{\langle \hat{n}(t_k) \rangle} = \frac{1}{2} .
\end{equation}
A statistics of this type corresponds to a noise level half the
shot-noise limit and is a signature for amplitude squeezing.  In fact,
the solution of Eq.~(\ref{eq:binom-recurrence}) can be given as a sum
of Binomial distributions, each with vibrational mean number
$l\!+\!k/2$ and variance $k/4$, that are weighted by the initial
number statistics $P_l(t_0)$:
\begin{displaymath}
  P_n(t_k) = \sum_{l=0}^n {k \choose n \!-\! l} \left( \frac{1}{2}
  \right)^k P_l(t_0) .
\end{displaymath}
Near the trapping state $|n_0\rangle$, however, the relative variance
deviates from the stationary value~(\ref{eq:stat-var}), decreases
below $1/2$ and reaches zero when the trapping state is finally
populated with unit probability.

Clearly, this behaviour is idealised by the fact that spontaneous
recoil effects during the periods of optical pumping have been
neglected in the analysis. However, using quantum-trajectory methods
we have included these effects consistently to obtain numerical
results. These results, that will be shown in the following, are based
on the parameters $\gamma_1 / \nu \!=\! 9.5$, $\gamma_2 / \nu \!=\!
3.3$, $\eta_1 \!\approx\! \eta_2 \!=\! 0.142$, and $|\Omega_{\rm p}|^2
/ \gamma_1^2 \!=\! 5.0$ for the optical pumping processes.

\begin{figure}
  \begin{center}
    \epsfig{file=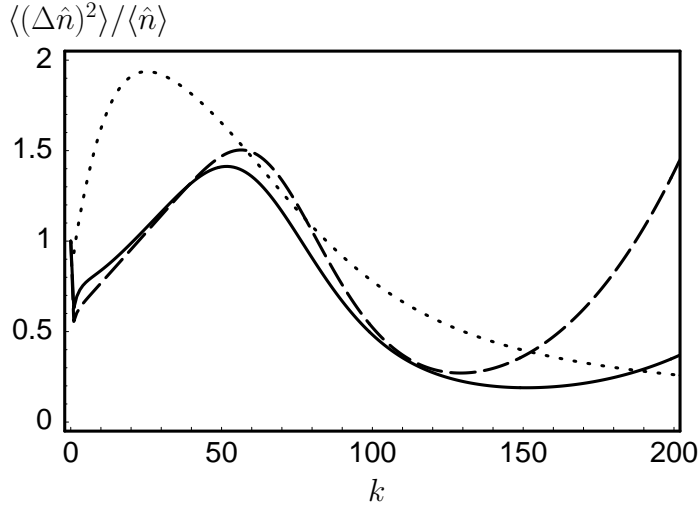,scale=0.7}
    \caption{Relative variance versus the number of cycles $k$ for the
      incoherent trapping state $|n_0\!=\!50\rangle$ with $\eta \!=\!
      0.268$. Spontaneous recoil effects during the optical pumping
      are included; saturation parameter is $\gamma s \tau_{\rm s}
      \!=\! 0.2 \mbox{ (dotted)}, 1.0 \mbox{ (solid)}, 2.0 \mbox{
        (dashed)}$.}
    \label{fig:relvar}
  \end{center}
\end{figure}

In Fig.~\ref{fig:relvar} the numerical results for the relative
variance of vibrational quanta is shown in dependence on the number of
cycles for the incoherent trapping state $|n_0 \!=\! 50\rangle$ for
varying effective saturations $\gamma s \tau_s$. It can be observed
that for the larger values of $\gamma s \tau_s$ the relative variance
transiently tends to the value $0.5$, but then increases due to the
photon scattering until the trapping state is approached. There the
relative variance decreases even below $0.5$, but only as a local
minimum, to monotonically increase afterwards. Furthermore, for low
``saturation'' (see dotted curve) the Binomial regime is not even
reached as an initial transient behaviour.

These results show that due to the recoil effects in the optical
pumping the trapping states are not truly stationary. Instead the
vibrational populations cross the trapping state to later be partly
accumulated at approximately equidistant further trapping states at
$n_0 \!\approx\! 100, 150, 200, \ldots$. However, it should be kept in
mind that the relative variance is very sensitive to minimal
populations at high vibrational quantum numbers. Thus a low percentage
of population distributed around or higher than the trapping state
drastically increases the relative variance. In this sense a more
suitable property is the maximum population reached in the trapping
state.   

\begin{figure}
  \begin{center}
    \epsfig{file=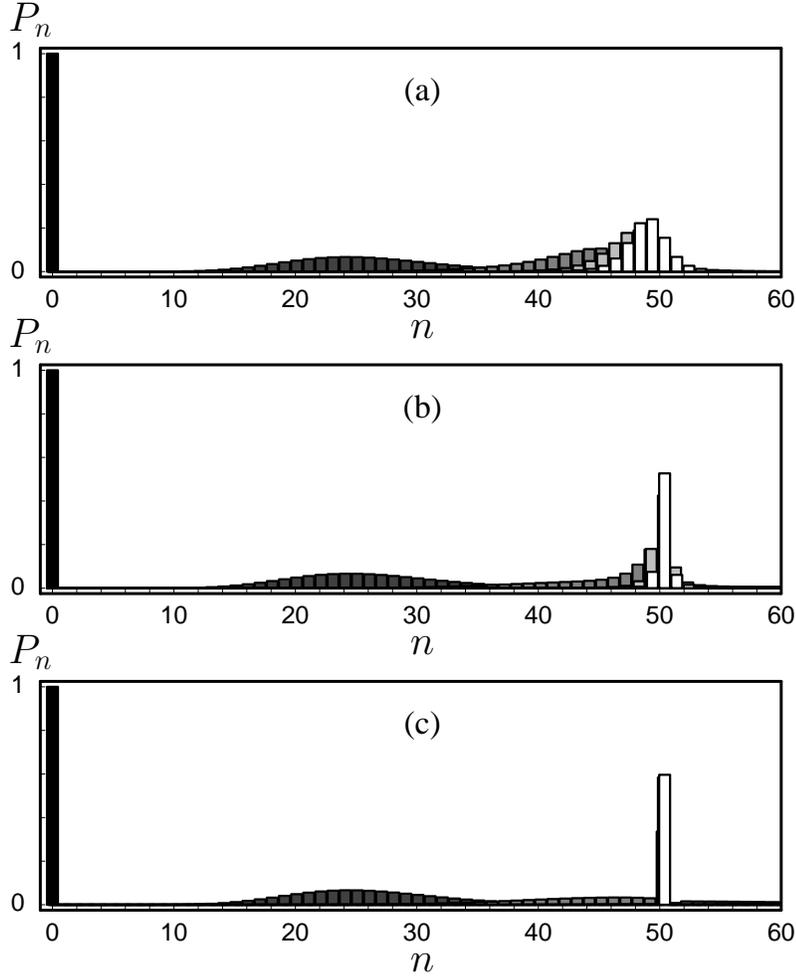,scale=0.9}
    \caption{Incoherent dynamics, including spontaneous recoil
      effects, of vibrational populations for $k \!=\! 0, 50, 100,
      150, 200$ cycles (gray scales from black to white) and
      Lamb--Dicke parameter $\eta \!=\! 0.268$ with trapping state
      $|n_0 \!=\! 50\rangle$. Other parameters are $\gamma s \tau_s
      \!=\! 1 \mbox{ (a)}, 10 \mbox{ (b)}, 1000 \mbox{ (c)}$.}
    \label{fig:incnl-high}
  \end{center}
\end{figure}

An example is shown in Fig.~\ref{fig:incnl-high} where vibrational
statistics is shown at different stages of the evolution for the same
trapping state $|50\rangle$. In fact it can be seen that the higher
the value of $\gamma s \tau_s$ the higher the population in the
trapping state at cycle $k \!=\! 200$. Clearly in the further time
evolution this population decreases again by scattering events that
eventually distribute the population at higher vibrational levels.
Nevertheless, if the laser interactions are stopped appropriately when
a high population in the trapping state is reached, maximum
populations up to $60\%$ at the level $n_0 \!=\! 50$ can be obtained.

\begin{figure}
  \begin{center}
    \epsfig{file=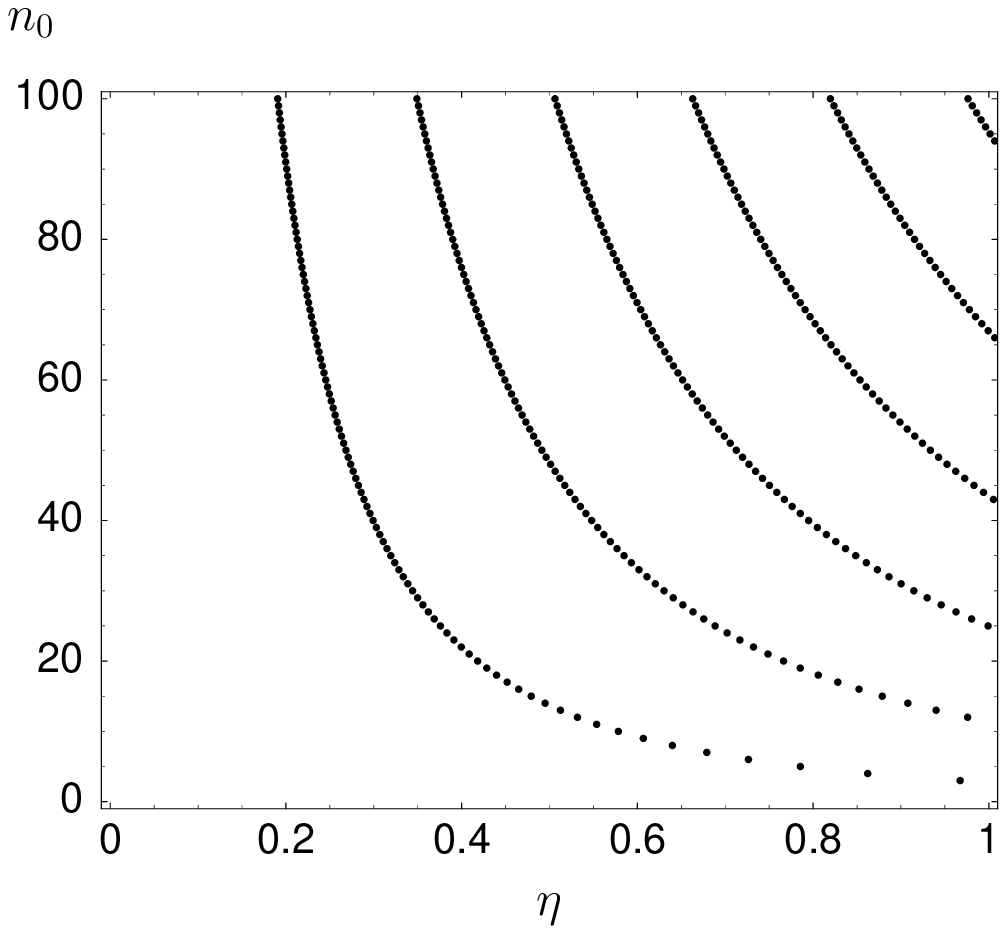,scale=0.8}
    \caption{Pairs of values for the trapping state number $n_0$ and
      the Lamb--Dicke parameter $\eta$ that obey the nonlinear
      trapping-state condition [Eq.~(\ref{eq:cond-nl})].}
    \label{fig:zeros}
  \end{center}
\end{figure}

It is noteworthy that for decreasing Lamb--Dicke parameters the
trapping states have increasing energies. In Fig.~\ref{fig:zeros} we
show certain combinations of Lamb--Dicke parameters and vibrational
quantum numbers $n_0$, that satisfy trapping. These are directly
obtained from condition~(\ref{eq:cond-inc}), together with
Eq.~(\ref{eq:gamma-laguerre}).

As to practical implementation of the incoherent method, experiments
on a single ion have been performed that indicate breakdown of the
ion-laser coupling at small values of the Lamb--Dicke parameter and
rather highly excited trap levels~\cite{toschek1,toschek2}. In these
experiments certain metastable states of the ion's vibrational
amplitude with $5 \mbox{--} 30\mu{\rm m}$ spatial extension within the
$400\mu{\rm m}$-wide focus of saturating laser light have been
observed. In steady state, one of these metastable states was found
occupied by the unperturbed, moderately laser-cooled ion. The ion
jumped to another one upon external perturbation: small variation of
the cooling power by minute laser detuning, irradiation by a weak
pulse of resonant radio frequency ($1 \mbox{--} 2$MHz, typically),
etc.  Each stable amplitude corresponds to a number of mean
vibrational excitation that is associated with a particular level, a
trapping level, where the laser-ion coupling breaks down for the given
small Lamb--Dicke parameter. Thus the nonlinearity in the laser-ion
coupling as discussed here, is of great importance even with a small
Lamb--Dicke parameter $\eta$ that makes trapping states appear at very
large vibrational quantum numbers.  These observations demonstrate
that an individual ion confined in a harmonic potential well indeed
may be prepared in a highly excited incoherent trapping state.

\section{Summary and conclusions} \label{sec:6}
We have studied ways of generating various kinds of ``trapping
states'' of the quantised centre-of-mass motion of a trapped ion. The
first kind is of the micro-maser type. In this case, the ion, at a
particular vibrational quantum number, undergoes a complete Rabi cycle
of excitation, such that higher quantum states cannot become excited.
This mechanism, however, is highly fragile with respect to any kind of
decoherence.  Moreover, when one intends to prepare a certain
vibrational number state, quasi-trapping conditions may exist for
other (lower) number states.  This feature leads to a substantial
increase of the time needed for the preparation.

Another way of implementing a trapping-state condition is based on the
nonlinearities that appear in the dependence of the atom-field
coupling on the vibrational excitation. These nonlinearities arise as
Franck--Condon-type overlap integrals due to the exchange of momentum
between the ion's centre-of-mass motion and the absorbed and emitted
laser photons. At certain vibrational quantum numbers the nonlinearly
modified interaction strength vanishes, leading to the expected
trapping states in a way quite different from the trapping mechanism
based on the completion of Rabi cycles. This trapping mechanism, when
used in its coherent version, reduces the preparation time, although
this method is still highly sensitive with respect to decoherence.

The particular advantage of the nonlinearity-based trapping condition
consists in the vanishing coupling for certain vibrational
excitations. This feature can be exploited for the generation of
trapping states even in situations where substantial decoherence
prevails, since now the trapping state is solely determined by the
selected Lamb--Dicke parameter. Moreover, the transition rates for
approaching the trapping state are more or less independent of the
vibrational excitation, which leads to substantial decrease of the
time needed to prepare a desired trapping state. Thus, one avoids the
sensitivity to decoherence, especially when preparing vibrational
number states of large quantum numbers.

\ack
This research was supported by Deutsche Forschungsgemeinschaft.


\begin{thebibliography}{99}
\bibitem{comment} Note, that level $|3\rangle$ in Fig.~1a, used for
  optical pumping, can be different from the level $|3\rangle$ in
  Fig.~1b, used for Raman driving the transition $|1\rangle
  \!\leftrightarrow\! |2\rangle$, to allow for independent control of
  Lamb--Dicke parameters.
\end{thebibliography}
\end{document}